\documentclass[a4paper]{article}

\usepackage{INTERSPEECH2020}
\usepackage{hyperref}

\title{Laughter Synthesis: \\Combining Seq2seq modeling with Transfer Learning}

\name{No\'e Tits~\href{https://orcid.org/0000-0002-1971-4412}{\includegraphics[scale=0.7]{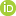}}, Kevin El Haddad~\href{https://orcid.org/0000-0003-1465-6273}{\includegraphics[scale=0.7]{orcid.png}}, Thierry Dutoit}
\address{
  Numediart Institute, University of Mons
  }
\email{\{noe.tits, kevin.elhaddad, thierry.dutoit\}@umons.ac.be }

\begin{document}

\maketitle
\begin{abstract}


Despite the growing interest for expressive speech synthesis, synthesis of nonverbal expressions is an under-explored area. In this paper we propose an audio laughter synthesis system based on a sequence-to-sequence TTS synthesis system. We leverage transfer learning by training a deep learning model to learn to generate both speech and laughs from annotations.
We evaluate our model with a listening test, comparing its performance to an HMM-based laughter synthesis one and assess that it reaches higher perceived naturalness.
Our solution is a first step towards a TTS system that would be able to synthesize speech with a control on amusement level with laughter integration.

\end{abstract}
\noindent\textbf{Index Terms}: Laugh synthesis, speech synthesis, TTS, deep learning, transfer learning


\section{Introduction and Motivations}


Given the progress in speech technologies and Human-Agent Interactions~(HAI), several applications of voice assistants and virtual agents have been developed. These applications are evolving towards breaking the barriers between robot-sounding synthetic sounds to human-like conversations. One of the under-explored domains is the synthesis of nonverbal conversational expressions, particularly laughter. Laughter is an important component of speech and daily interactions. It has been shown to be very frequent cross-cultural expression in conversations, to communicate emotions and to have conversational and social functionalities~\cite{laskowski2007analysis,devillers2007positive,soury2014smile}.

Laughter can be expressed in many different ways and is particular to each individual. It is therefore rather difficult to collect naturalistic genuine laughter in a sound clean environment. This is the main reason why resources available for synthesis purposes are rather limited compared to speech.

In this paper we present a deep learning-based laughter synthesis system. This work is part of a larger project aiming at synthesizing laughter alongside speech in HAI systems. 
An initial system was trained from scratch using the same deep-learning based model described here and only laughter data. The results obtained were not satisfying. This was probably due to the limited amount of data available for training. It thus motivated the use of speech data as well. Indeed, although different, both speech and laughs share common sound characteristics since laughs are sequences of fricatives, vowel-sounds and breathing. So, we leverage the knowledge learned by TTS systems when trained with speech in order to improve laughter synthesis. 

The paper is organized as follows: 
related work is summarized in Section~\ref{related_work_laughter}; 
Section~\ref{laughter_dataset} presents the datasets involved in this work;
Section~\ref{system_description_laughter} describes the proposed system for audio laughter synthesis;
the procedure of the perceptive evaluation is described in
Section~\ref{evaluation_laughter}; 
the results of the evaluation are presented and discussed in Section~\ref{results_laughter}; 
finally we conclude and detail our plans for future work in Section~\ref{conclusions_laughter}. Most of the resources used here are accessible~\footnote{https://github.com/numediart/LaughterSynthesis}.

\begin{figure*}[h]
  \centering
  \includegraphics[width=1\textwidth
  ]{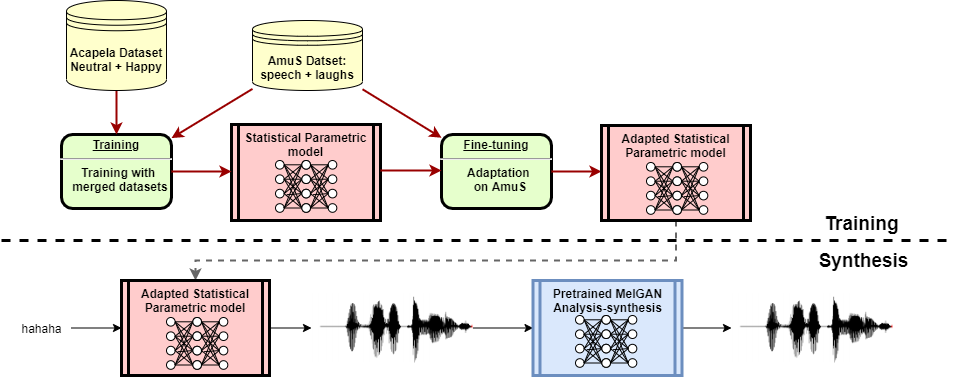}
  \caption{Block diagram of the proposed method for model adaptation. The statistical parametric model is based on DCTTS and described in Section~\ref{laughter_system_description}. The text input showed on the figure is symbolic, its detailed format is described in Section~\ref{laughter_dataset}.}
\label{laughter_synthesis_block_diagram}
\end{figure*}

\section{Related Work}
\label{related_work_laughter}


The techniques studied for laughter synthesis have generally followed those used for speech synthesis. In this work we will do the same for two main reasons: first, the signals share a lot of common characteristics; second, using TTS systems will allow to incorporate our laughter synthesis system into a fully functioning TTS system and thus achieve our ultimate goal of TTS with control over amusement levels.

Speech synthesis methods can be grouped in three main categories: synthesis by concatenation, parametric synthesis and statistical parametric synthesis~\cite{theory_behind_control_tts-19-tits}.
Among the few studies on laughter synthesis, the first attempts included techniques like synthesis by diphone concatenation~\cite{laughter_articulatory-07-trouvain}, parametric synthesis and by using a mass-spring approach~\cite{laughter_parametric_lpc-07-sundaram}.
Then Hidden Markov Models (HMM)-based models were introduced to laughter synthesis due to their wide-use in speech synthesis back then~\cite{HMM_laughs-13_urbain}.

In our previous work we used an HMM-based approach to synthesize laughter alongside amused speech in~\cite{kev15speechLaugh}. Synthetic laughs were obtained by training HMM systems on laughter data. The amused speech was obtained by adapting HMMs trained with another speaker's neutral speech data and adapting them to a smaller dataset containing speech from the speaker from which the laughs were recorded (due to the limited amount of data available for that speaker).

However, deep learning-based laughter synthesis has been little explored yet. A recent approach of synthesis with wavenet was recently proposed by \cite{laughter_synthesis_interspeech-19-mori}. Wavenet~\cite{wavenet-16-vandenoord} is an autoregressive CNN synthesizing audio sample by sample from features, typically linguistic features for TTS, or acoustic features for vocoding.
In their attempt of application to laughter synthesis, they conditioned the wavenet model on information of inhalation/exhalation sequence and parameters of durations and power contour predicted by an HMM model. This approach is therefore still relying on previous HMM approaches for a part of the information.

Given the breakthrough of sequence-to-sequence (seq2seq) approaches in speech synthesis systems, we propose an adapted approach for audio laughter synthesis.
Along with the synthesis quality, the method proposed in this paper also offers control over the specific sound sequences to be generated rather than syllable-level control. This allows the flexibility of choosing the specific sequence of voiced (vowels) and unvoiced sounds to be generated.
Given the aforementioned goal of obtaining a fully functioning TTS system generating laughter alongside speech, another advantage is not only to synthesize naturalistic human-like laughs, but also to do it in a speech context. \\\\
This will allow a later integration in a fully functioning speech and laugh synthesis system as planned.

\section{Dataset}
\label{laughter_dataset}
In order to apply the transfer learning approach described above, the data used are formed of subsets of a proprietary dataset recorded by Acapela and of the AmuS dataset~\cite{AmuS17emoDB}.

Acapela's dataset was recorded to build a narrating framework to construct book recordings from transcriptions.
It contains phonetically rich sentences uttered by a male actor in US English. The actor was asked to utter a set of the sentences in 8 style classes.
For the purpose of this work, only the audio recordings of the neutral style were kept along of the corresponding transcription with a total of 150.50 minutes (3299 utterances) of speech data.

The AmuS dataset contains recordings of amused speech components such as smiled speech, laughs and speech-laughs. For this work, we chose the speaker with the most amount of recorded laughs: SpkB.
In this dataset, the purpose of the laughs were to be inserted in speech in order to create an amused effect. This suits well with the goal of generating laughs alongside speech as mentioned above.

In order to record these, the subject was asked to watch stimuli of funny content while sustaining the sound of a vowel, until eventually laughter occurred, naturally interrupting the vowel. This would allow us to collect laughs with transitions from and to vowels. We thus have at our disposal laughs occurring in three vowel contexts: [a], [e], [i] (French IPA symbols). Table~\ref{tab:laughs_count} breaks down amount of data available.

 \begin{table}[]
     \centering
     \begin{tabular}{|c|c|c|c|c|}
\hline
Vowel             & {[}a{]} & {[}e{]} & {[}i{]} & Total \\
\hline
number of samples & 54      & 33      & 25      & 112   \\
\hline
duration (sec) &  101       & 63         &  38       & 202     \\ 
\hline
     \end{tabular}
     \caption{Quantity of laughs per vowel context.}
     \label{tab:laughs_count}
 \end{table}
 
 Isolated laughs are sequences of voiced and unvoiced sounds. In AmuS, the laughs were segmented and each segment was given a label corresponding to a voiced or unvoiced category. 
These label sequences and the corresponding laughter audio signals were used to train our systems.
 


\section{Seq2seq Audio Laughter synthesis}
\label{system_description_laughter}

\subsection{System description}

\label{laughter_system_description}

Nowadays, one of the major techniques for Text-to-Speech synthesis are deep learning architectures based on the sequence-to-sequence (seq2seq) principle.
It consists of an encoder-decoder setup with an interface between the two components called \textit{Attention Mechanism} whose role is to model the alignment between input and output sequences. Well known seq2seq TTS systems are Tacotron~\cite{tacotron-17-wang}, Char2wav~\cite{char2wav-17-sotelo} and DCTTS~\cite{dctts-17-tachibana}. In this work we adapted the DCTTS model for audio laughter synthesis using an open implementation available online~\footnote{https://github.com/CSTR-Edinburgh/ophelia}.

In this work, the input sequence is composed of speech phonemes and laughter annotations described in Section~\ref{laughter_dataset}.

We use festival~\cite{festival-97-black} to extract phones from transcriptions in the Acapela dataset and the speech part of AmuS dataset.

The output sequence is a mel-spectrogram.
A second part of DCTTS, trained separately reconstructs a full resolution magnitude spectrogram from the mel-spectrogram to be inverted to a waveform using Griffin-Lim algorithm~\cite{griffin_lim-84-griffin}. For more details about these, see~\cite{dctts-17-tachibana}.

A first Text-To-Speech and Laughter is trained using the speech from the Acapela dataset due to the quantity of data it provides along with the laughs and smiled speech from the speaker SpkB of the AmuS dataset.
This system is then fine-tuned with smiled speech and laughter, both coming from the same AmuS speaker (SpkB). This was done  with the perspective of integrating this system into a fully functioning TTS system with control over amusement, as mentioned previously.

This adaptation technique was previously tested on emotional speech and showed promising results in~\cite{exploring_transfer_learning-19-tits}, which motivated its use in this work.

Figure~\ref{laughter_synthesis_block_diagram} shows a block diagram of the procedure proposed for model adaptation and waveform correction with MelGAN.


\subsection{Waveform Correction with MelGAN}

To generate the waveform from acoustic features, it has been shown that neural audio synthesizers achieve better quality in terms of naturalness~\cite{wavenet-16-vandenoord, wavernn-18-Kalchbrenner}. However it is generally a challenge to design and optimize such models efficiently to reach the expected results described in the literature.
They also often lose generalization properties compared to signal processing based vocoders, as they are often speaker dependent.

MelGAN~\cite{melgan-19-kumar} is a recently proposed model that tackled the problems of efficiency and generalization accross speakers. The model is non-autoregressive, fully convolutional and smaller than previous ones. 

In this paper we use MelGAN as a waveform corrector. The laughter waveform is first synthesized by the system described in Section~\ref{laughter_system_description}. This waveform contains artifacts due to the Griffin-Lim estimation. Then we apply analysis and synthesis with MelGAN to obtain a corrected laughter waveform that, we show, is perceived as more natural compared to not using MelGAN.

\section{Perception Test}
\label{evaluation_laughter}



To evaluate the obtained results we set-up a perception test using a Mean Opinion Score~(MOS) test.
We gathered the samples from these different methods:

\begin{itemize}
    \item Method 1: original laughter samples from AmuS dataset (SpkB) which will serve as a top-line of naturalness for the different methods.
    \item Method 2: Synthesized laughs based on the HTS system equivalent to the one used in~\cite{kev15speechLaugh}
    \item Method 3: 
    seq2seq model (seq2seq-GL) described in Section~\ref{laughter_system_description}.
    \item Method 4: same seq2seq model as method 3 followed by the proposed MelGAN waveform correction (seq2seq-MelGAN)
\end{itemize}


The MOS test focused on evaluating the naturalness of the synthesized samples of the methods. It was implemented as a web experiment with turkle\footnote{https://github.com/hltcoe/turkle},  which is an open-source web server with which one can host a crowdsourcing application locally,

A total of 71 samples were presented to 24 participants in a random order (the participants characteristics are detailed in Table~\ref{tab:n_participants})). 
They were asked to rate each sample in terms of naturalness on a 5-point Likert scale with the
following labels: very unnatural (score 1), unnatural (2), fairly natural (3), natural (4) and  very natural (5). 
They could listen to each sample as much times as needed and could stop at any point of the test.

The definition of naturalness for speech and laughter could be interpreted in different ways during evaluation. For example it would have been possible that  a participant would rate the acting quality instead of human-likeness of the sound perceived during the listening test. Indeed, "not natural" can be perceived as "fake" or "simulated" instead of "synthetic". But in order to not deviate from previous work and evaluations that used the word "natural", we preferred using it and specify what we mean by it.

That is why, we added an explanation of the meaning of "natural" in the question asked, focusing on the definition of "human-likeness" for "natural".

We also asked some of the participants at the end of the test to comment on what mainly influenced their choices. This mainly serves as our qualitative testing.



\begin{table}[]
\centering
\begin{tabular}{|c|c|c|c|c|}
\hline
             & Female & Male & Sum \\
\hline
[20,40[ &    7   & 13         & 20   \\
\hline
[50,65[ &  1      &  3       & 4     \\ 
\hline
Sum &  8      &  16       & 24     \\ 
\hline
\end{tabular}
\caption{Number of participants by gender and age range (in years)}
\label{tab:n_participants}
\end{table}

\section{Results}
\label{results_laughter}

In this section we present the results obtained from the perception test. These are divided in quantitative and qualitative analyses. The former presents the scores obtained along with interpretations while the latter reports on comments made by the participants on what influenced their choices,

\subsection{Quantitative Analysis}
A total of 1696 answers were collected. Table~\ref{tab:MOS_table} gathers the number of ratings for each method, the resulting MOS scores and their standard deviation.

\begin{table}[]
\centering
\begin{tabular}{|c|c|c|c|c|}
\hline
{} &  \#ratings &   MOS &   std \\
\hline
HMM            &     407 &  2.64 &  1.02 \\
seq2seq-GL     &     431 &  2.50 &  1.09 \\
seq2seq-melgan &     429 &  3.28 &  1.06 \\
original       &     429 &  4.10 &  0.91 \\
\hline
\end{tabular}
\caption{Number of collected ratings, MOS scores and their standard deviation for each method}
\label{tab:MOS_table}
\end{table}

\begin{figure}[h]
  \centering
  \includegraphics[width=1.1\linewidth
  ]{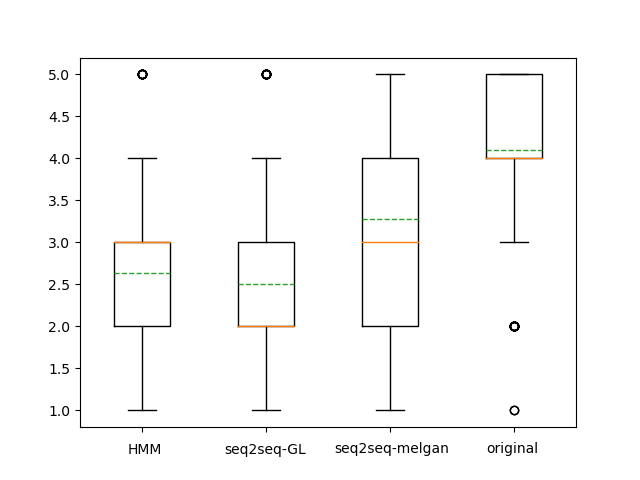}
  \caption{Boxplots of scores distributions of the different methods. The green lines correspond to the Mean Opinion Scores while the red lines show the median values of the scores.}
\label{laughter_boxplot}
\end{figure}

\begin{figure}[h]
  \centering
  \includegraphics[width=1.1\linewidth
  ]{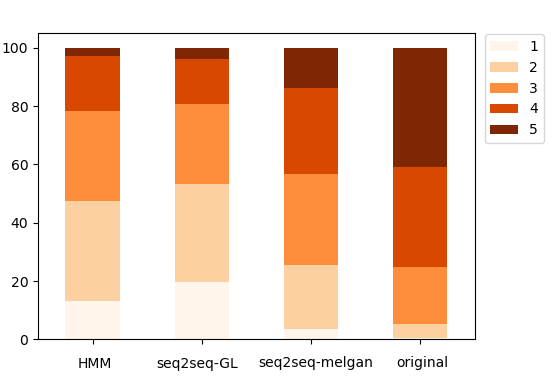}
  \caption{Score distributions of the different methods.\\ The scale ranges from 1:very unnatural to 5: very natural.}
\label{laughter_distribution}
\end{figure}

Figure~\ref{laughter_boxplot} shows the distributions of the ratings as boxplots. Figure~\ref{laughter_distribution} shows the percentage of scores chosen for each method.

Original samples of the dataset reach a MOS score that is below 5 as listeners do not rate all original samples as perfect. There is a quite high variance, that is close to one for all methods, which was also the case in~\cite{HMM_laughs-13_urbain}. 

We can clearly see from the results obtained that our seq2seq-MelGAN outperformed both the seq2seq-GL and the HMM-based systems. Both seq2seq-GL and HMM-based system obtained similar MOS.
 We interpret these results by attributing the loss in MOS obtained, in the latter 2 systems generated laughs, to the distortions found in the seq2seq-GL laughs and the more robotic effect found in the HMMs-obtained laughs.
We base our interpretation on previously reported results as detailed in what follows, but also on our personal observations that were not officially tested yet and ones reported by the MOS tests participants as detailed in Section~\ref{sec:qual-anal},

This shows that the seq2seq is efficient at synthesizing laughter, but the question asked in the MOS test to grade naturalness (human-likeness) combined with the distortion generated by the Griffin-Lim algorithm degraded the grades obtained for this test.


Although MOS tests are never executed exactly the same way, it is always interesting to analyze and compare the results obtained here with respect to the results of previous similar experiments found in the literature. This will also partly back our results interpretation.

In~\cite{melgan-19-kumar}, the authors compared MelGAN vocoder to the Griffin-Lim algorithm for seq2seq TTS. They reported the Griffin-Lim algorithm to be responsible for a large part of the distortion leading to a loss of 2.95 MOS points compared to original samples.
MelGAN, on the other hand, offered a gain of 1.77 points of MOS over the Griffin-Lim algorithm. In this work, the MelGAN waveform correction offered a gain of 0.78 points of MOS. It is important to highlight the fact that training MelGAN directly on generated spectrograms will likely improve our results which was not done in this study but will be part of future work.
In~\cite{HMM_laughs-13_urbain}, the authors compared different variants of HMM-based laughter synthesis. They show that the distortion caused by the vocoder in the copy-synthesis samples is of 0.8 compared to original samples. Their best synthesis solution is 0.6 points below that, and therefore 1.4 below original samples. In this paper, the HMM approach is 1.46 below original samples which is close to their results.
In~\cite{laughter_vocoder-14-urbain}, they confirm that HMM-based laughter synthesis have significantly lower quality than copy-synthesis with several vocoders.






\subsection{Qualitative Analysis}
\label{sec:qual-anal}
A part of the participants were asked to comment on what, according to them, influenced their choices in rating the laughs during the test. In this section, we summarise the obtained qualitative results. We report these comments in order to shed a bit more light on the results obtained from the quantitative test. Here is a list of the what some of the participants reported:

\begin{itemize}
\item The duration seemed to be another important parameter to consider, as some participants seemed to have based their choice on it. Indeed they reported that some laughs were too long/short to be natural.
Also, some participants found short laughs to be ambiguous on a naturalness scale, as if they were focused on finding the fake laughs from the real ones instead of solely focusing grading how natural they perceive the laugh they were listening to.
\item Laughs with varying pitch and duration seemed to be perceived as more natural as opposed to laughs with a monotonous prosody. Some laughs were described as "sounding like a repeated sequence" and were perceived as robotic (these would correspond to the HMM generated laughs) whereas laughs containing more randomness were perceived as more natural. Similarly, the ones of which the loudness was fading off (decreasing until the end) were perceived as more natural than ones with a monotonous loudness level and ending abruptly.
\end{itemize}

Apart from the main parameters influencing degrading decision during the MOS test, we note how important the choice of the question and phrasing is for subjective evaluations such as these.

\section{Conclusions and Future Work}
\label{conclusions_laughter}

In this paper, a new approach of audio laughter synthesis based on seq2seq learning was proposed inspired by the evolution of the TTS field. This system is implemented by leveraging the patterns learned to pass from text to acoustic features in speech, to learn laughter synthesis.

We also use a pretrained MelGAN model as a post waveform corrector allows to remove audio artifacts generated by Griffin-Lim algorithm and thus improve the scores obtained in a MOS test.
We believe several modifications could improve the acoustic quality of the synthesis. First end-to-end training could help concerning the accumulation of errors of several blocks: the seq2seq system and the vocoder.

This results in a strong improvement over past methods of audio laughter synthesis (including our own \cite{kev15speechLaugh}) in terms of naturalness and is promising for later use to build amused speech synthesis systems.
The promising results obtained here, allows us to work on incorporating the laughter synthesis system into a fully functioning TTS with control over amusement level. The fact that our laughter synthesis system was developed in a TTS context makes this integration easier.

\section{Acknowledgements}

No\'e Tits  is funded through a FRIA grant (Fonds pour la Formation \`a la Recherche dans l'Industrie et l'Agriculture, Belgium). 

\pagebreak

\bibliographystyle{IEEEtran}

\bibliography{mybib}
\end{document}